\documentclass[10pt]{article}
\usepackage{multicol}
\usepackage{comment}
\usepackage{wrapfig}
\usepackage{graphicx}

\usepackage{graphicx, wrapfig, subcaption, setspace, booktabs, float}
\usepackage[T1]{fontenc}
\usepackage{lipsum}
\usepackage[footnote,printonlyused]{acronym}
\usepackage{scrextend}
\deffootnote{0em}{1.6em}{\thefootnotemark.\enskip}

\usepackage[font=small, labelfont=bf]{caption}
\usepackage{fourier}
\usepackage[protrusion=true, expansion=true]{microtype}
\usepackage{indentfirst}
\usepackage{textcomp}
\usepackage[table]{xcolor}
\usepackage{enumitem} 
\usepackage{ragged2e} 

\usepackage{xcolor}
\usepackage{soul}

\usepackage{hanging}

\usepackage[english]{babel}
\usepackage{csquotes}

\usepackage[a4paper,top=2cm,bottom=2cm,left=2cm,right=2cm,marginparwidth=1.5cm]{geometry}
\begin{document}
\pagestyle{plain}
\setstretch{1.2}

\pagenumbering{gobble}

\vspace*{5cm}

\begin{center}
    {\Huge \textbf{An Outline for a Jupyter-Materials-Based Repository Website Focused on the Computational Sciences}}
\end{center}

\vspace*{3cm}

\begin{center}
    {\Large  Authors}\\
    \vspace*{1cm}
    Zachary Kelly \\
    Augustana Campus, University of Alberta\\
    4901 46 Ave, Camrose AB T4V 2R3, Canada\\
    \vspace*{1cm}
    Peter Berg \\
    Faculty of Mathematics and Science, Brock University\\
    1812 Sir Isaac Brock Way, St. Catharines ON L2S 3A1, Canada
\end{center}

\newpage
\pagenumbering{arabic}

\begin{center}
    {\Large \textbf{An Outline for a Jupyter-Materials-Based Repository Website Focused on the Computational Sciences}}
\end{center}

\begin{multicols}{2}

\textbf{\textit{Abstract}---As access to the internet has become increasingly ubiquitous, along with the reliability and speed of internet providers, so too has the implementation of internet-based learning tools. These tools provide students opportunities to do meaningful work away from university, however, often at a financial cost to universities and students. Moreover, limited and high-cost internet access in less-developed countries and remote areas acts as a barrier to implementing these tools in a meaningful way, leading to inequalities in both the quality of education and the opportunities provided. This paper outlines the development process, and benefits, of a low-cost and light-weight repository website centred around disseminating open-source textbooks and other supplemental learning materials for computational sciences using Jupyter Notebooks. The website focuses on allowing students to download their textbooks and other materials from a centralised location, to be used offline or with limited internet access. Internet access is not the only constraining factor; access to reasonably priced personal computers also limits the effectiveness of internet-based learning tools. As such, this paper will also explore the feasibility of integrating low-cost Raspberry Pi kits into this development process as a way of increasing the reach of an online repository of open-source Jupyter Notebook textbooks. While this paper focuses on Canadian universities and remote communities, many of the website's proposed applications are relevant worldwide.} 


\begin{center}
    {\Large \textbf{1.0 Introduction}}
\end{center}

Since the emergence of COVID-19, online and distance learning has become an area of focus and development in universities where such methods were not required previously, creating opportunities for the introduction of new tools and pedagogical methods into teaching (Mushtaha et al., 2022). Notably, Project Jupyter (\textit{Project Jupyter}, n.d.-a) has been shown to provide tools that allow interactivity and convenience for both students and instructors when integrated into computational-science based distance learning (Noprianto et al., 2022). \\

As universities and students are recognizing the impact that open-source textbooks and materials have on both learning outcomes and cost savings (Fischer et al., 2015), many universities have begun to provide open-source educational resources for their students, including the University of Waterloo and the University of Toronto (Open UToronto, 2023; \textit{Open Educational Resources}, n.d.). While most open-source textbooks available are in eBook or PDF form, Jupyter Notebooks\footnote{Henceforth referred to as JNB}(\textit{Project Jupyter Documentation}, n.d.), a browser-based interactive computing environment, are capable of delivering computational-sciences textbooks that provide a synergistic work flow between theoretical text and its practical application through JNB's code cells (Weiss, 2020). Jupyter Books\footnote{Henceforth referred to as JB}, an extension of JNBs are a compiled directory of JNBs and Markdown files\footnote{Markdown is a lightweight and simple text-to-html syntax that is used to format text on web pages (\textit{Daring Fireball: Markdown}, n.d.)} that allow for a browser-based, navigable interface persevering the functionality of JNBs (\textit{The Executable Books Project}, n.d.). We will refer to the collection and use of both JNBs and JBs as Jupyter materials, or JMs.\\

Furthermore, due to JM's open-source nature and their capacity to function both online and offline, JM-based computational textbooks are well suited for areas of the world with poor internet infrastructure and access. In order to share and make use of these textbooks, we suggest that a JM-based repository website would benefit not only universities beginning to adopt distance learning methods but also universities where students have limited access to the internet and online learning tools. Such a website would be able to host JNBs, make them available to be viewed and downloaded, be as light-weight as possible in order to reduce bandwidth usage, and be free to use. \\

In conjunction with providing an open-source JM textbook platform, providing low-cost methods of utilizing their functionality is paramount. Recently we have seen the adoption of Chromebooks as a low-cost alternative to traditional laptops in primary and secondary education (Quin, 2020). Yet, Chromebooks often require some degree of circumvention with regards to compatibility issues for certain programming languages and programs. In contrast, Raspberry Pi (\textit{Raspberry Pi Documentation}, n.d.) provides a low-cost and Linux-based environment in which JNBs can operate, along with other educational programs such as Wolfram, Mathematica and Scratch (Kurniawan, 2018), while matching the processing power of many mid to low tier Chromebooks. \\

This paper will outline (1) the benefits and processes of using a light-weight JM based repository website, (2) the benefits of adopting open-source JM-based textbooks and supplementary materials for the computational sciences, (3) how Raspberry Pis can provide low-cost computational power in remote and fiscally-constrained communities, and (4) how the combination of these tools produce an environment which facilitates the transfer of expertise from community to community, strengthening the capabilities of individual students and teachers, while increasing the availability of knowledge and pedagogical methods. This outline is based on an ongoing project in partnership with the University of Eswatini, and Academics Without Borders, focused on delivering a JM-based repository website that can be used in conjunction with their existing computational-sciences curriculum.\\


\begin{center}
    {\Large \textbf{2.0 Project Jupyter}}
\end{center}

Project Jupyter is an open-source project focused on creating web-based interactive computing environments that allow users to write and execute code in a REPL (read-eval-print loop) manner (\textit{Project Jupyter}, n.d.-b). Currently, Project Jupyter is developing JNBs which are an open-source browser-based application that facilitates the creation of interactive journals, textbooks, and other documents that contain executable code, equations, text, and visualisations such as output from code, videos, and images (see Figure 1). Moreover, Project Jupyter has developed a web-based platform for working with JNBs called Jupyter Labs, which is a flexible and modular interface that can be utilized to easily build and edit JNBs (\textit{JupyterLab Documentation}, n.d.). \\

\begin{figure*}[!ht]
    \begin{center}
    \includegraphics[scale=0.6]{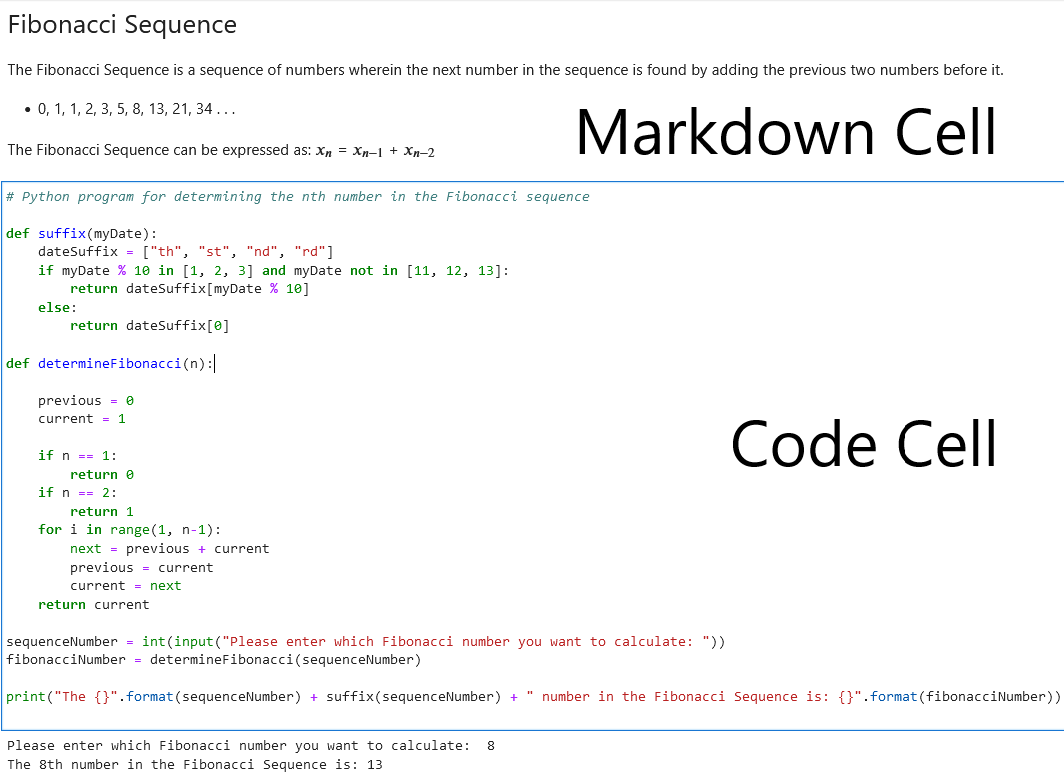}
    \captionof{figure}{Python Fibonacci Sequence Worksheet}
    \end{center}
\end{figure*}

JNBs natively support Python. Furthermore, there are currently over a hundred different kernels available for JNBs that can be installed to support over 40 different programming languages such as Scala, Julia, R, C/C++ and others (\textit{Kernels (Programming Languages)}, n.d.). These programming languages are written within JNB code cells, which are modularized containers that contain the code to be executed by the JNB's kernel; variables and other data are preserved between these containers. JNBs also support Markdown text resembling the LaTex environment; JNBs can be created containing no code cells at all, meaning that many documents that can be created as PDFs can also be created as JNBs. Users can convert these Markdown Notebooks, as well as those that contain code, into well formatted PDF using LaTex templates by utilizing nbconvert (\textit{Nbconvert: Convert Notebooks to Other Formats}, n.d.) which is currently included in the Jupyter environment. Output from code cells such as print statements, or images such as graphs can also be preserved in conversion, allowing for both a code cell and its output to be viewed statically in PDF and HTML files (\textit{Markdown Cells}, n.d.). \\

Project Jupyter has also been integrated into other open-source projects such as an extremely popular extension for VsCode (\textit{Working With Jupyter Notebooks in Visual Studio Code}, 2021) and in the base installation package for Anaconda (\textit{Jupyter Notebooks — Anaconda Documentation}, n.d.), simplifying the installation, creation, and ease of use for JNB. JNBs can also be executed on the Google Cloud by directing Google's Colab platform to Notebooks stored on GitHub (\textit{Google Colaboratory}, n.d.). The same effect can be achieved by directing Binder\footnote{Binder is a Jupyter adjacent project that allows for the reproduction, and sharing, of custom computing environments that can be used to operate JM's remotely from online repositories} to a repository of JNB, with the added functionality of being able to display several stored Notebooks concurrently (\textit{Binder Documentation}, n.d.). Traditionally, JNBs have been used in the data sciences, largely for their ability to document and explore data both in a linear and non-linear fashion (Perkel, 2018). However, other fields have adopted JNBs; most notably, the LIGO and Virgo collaboration released their data and research on the first direct observation of gravitational waves in 2016 as a JNB (\textit{Tutorials}, n.d.) (see Figure 2). \\ 

\begin{figure*}[ht]
    \begin{center}
    \includegraphics[scale=0.8]{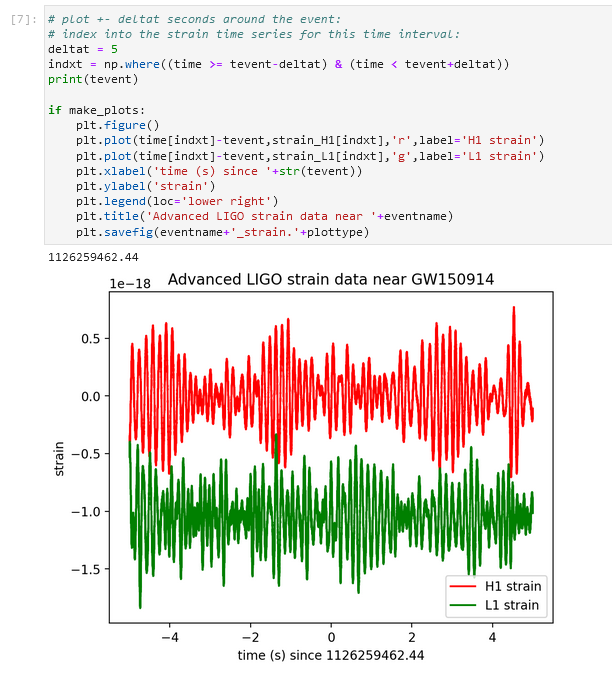}
    \captionof{figure}{Graphical Output from the LIGO Notebook Tutorial}
    \end{center}
\end{figure*}

Many researchers, instructors, and students prefer JNBs to traditional Python scripts for several reasons:

\begin{itemize}[leftmargin=.15in]
\item Minimal Complexity - To run JNBs, there is no need for a new Operating System nor is there any need to install and download large applications. Jupyter Labs is also quite lightweight\footnote{Lightweight referring to Jupyter Lab's small installation package, low memory footprint, and reasonable CPU usage} and can be assumed to run on any system.
\item Maximal Flexibility - Installing and configuring the packages needed for any project outside the scope of the Jupyter Labs/Notebooks base package installation is simple.
\item Interactivity - Explanations and code can easily be presented side-by-side, allowing for narratives to be expressed clearly, as well as allowing others to add their own annotations easily.
\item Utility - Provision of a ‘real-life’ development environment that represents a comfortable standard in both personal and professional situations.
\item Maintainability - On-going maintenance is provided by Project Jupyter, but easy updating  by instructors or project leads is also possible, allowing for version control and easy distribution (Reades, 2020; Ruiz-Sarmiento et al., 2021).
\end{itemize}

However, JNBs need not be used only as a supplementary tool for industry; their widespread use in the data sciences and other fields imply that there are avenues in which JNBs could find use in presenting information to secondary and post-secondary students. Due to JNB's built-in interactivity, one can develop textbooks and supplementary materials for computational subjects that go beyond what traditional materials in classrooms can deliver. It is important to note that while the initial installation of Jupyter, along with any required packages, requires an internet connection, the functionalities of JNBs can be operated without. This may not seem intuitive at first, as Jupyter Labs operates within an internet browser, but it is exactly one of the Jupyter Projects greatest strengths compared to many internet-based learning tools. The main benefits for students with regards to JNBs are as follows:

\begin{figure*}[ht]
    \begin{center}
    \includegraphics[width=\linewidth]{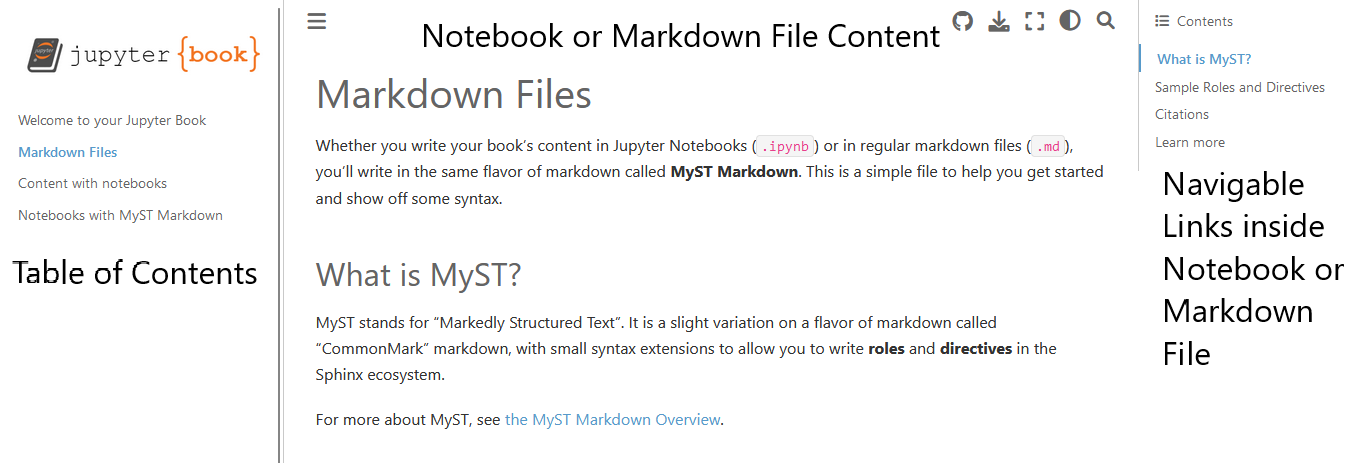}
    \captionof{figure}{Jupyter Book with Navigable Table of Contents Present}
    \end{center}
\end{figure*}
\begin{itemize}[leftmargin=.15in]

\item JNBs provide an avenue in which students can hone their ability to think computationally. Notebooks offer students an easy to see and manageable structure in which to decompose and break down problems into smaller sets. This decomposition creates the opportunity to observe and recognize any patterns, the space to identify the general principles that adhere to said recognized patterns, and, lastly, the ability for learners to use their skills to develop and recognize the steps required to solve comparable problems (\textit{Why We Use Jupyter Notebooks}, 2019).
\item Working in an environment that balances a requirement of higher-level concepts while still being user friendly. High-end commercial software may reduce computational insight into low-level concepts through its automation of certain tasks and requirements (\textit{Why We Use Jupyter Notebooks}, 2019).
\item The open-source nature of JNBs introduces students to the ever expanding set of open-source projects available today, while providing a low barrier of entry (\textit{Why We Use Jupyter Notebooks}, 2019).
\item Providing an interactive platform that instills active learning methodologies in the students. JNBs allow students to both consume and create/edit Notebooks as they progress in the course materials and the development of their skills (\textit{Why We Use Jupyter Notebooks}, 2019).
\end{itemize}

These added benefits that are derived from the use of JNBs have been shown to lead to a large and direct increase in the students' ability to learn course material compared to traditional textbooks. Both instructors and students have found value in “their utilisation, highlighting the quality of the resultant material, and its great contribution towards fluid lab sessions” (Ruiz-Sarmiento et al., 2021). JNBs have found educational use in teaching mathematics (Isihara, 2021), being used as part of a real-time spectrophotometer to enable spectroscopy during chemistry labs (Mitsioni et al., 2023), and assisting in the teaching and studying of advanced topics in fluid mechanics (Castilla \& Peña, 2023); owing to its versatility, there are opportunities to find uses for JNBs in almost every field of study. \\

Furthermore, due to JNBs' HTML based environment, their application in virtual and online learning environments allows for a much more seamless learning process compared to alternatives such as screen sharing static lab worksheets or coding demonstrations. When Covid-19 forced many institutions into on-line classroom scenarios, JNBs showed themselves to be quite useful in this new and adaptive environment, allowing for computational exercises to replace traditional lab exercises, even for those who had no previous coding experience (Pillay, 2020). Plugins have also been developed to aid and assist in the creation and grading of assignments based on JNBs, most notably nbgrader which allows instructors to create assignments in traditional notebook formats (\textit{Nbgrader}, n.d.). \\


\begin{center}
    {\large \textbf{2.1 Jupyter Books}}
\end{center}

Building on the functionality of JNB's and Project Jupyter is the Jupyter Book, an international collaboration focused on open source projects that aid and promote the publication of computational narratives using Project Jupyter programs (\textit{The Executable Books Project}, n.d.). JBs work by compiling, or building, JNBs and Markdown files into a viewable HTML site through JB's command line interface; this process preserves output from kernels containing code within the compiled JNBs as well as any formatting that the Markdown indicates (\textit{Build Your Book}, n.d.). This grants the user the ability to create navigable JBs containing the information and formatting from several files in a familiar browser-based ’chapter-esque’ format similar to a textbook such as in Figure 3. \\

JBs are not 'one file' documents like JNBs; in order to view the built HTML JB site, it is necessary that all HTML files that were generated be present, and all the paths between these HTML files are preserved. One can think of JBs as a local website where instead of HTML web pages being stored on a server and accessed through the Internet, they are stored locally on the user's machine. Thankfully, when creating a new JB, a directory is created to store and maintain the built HTML and paths required for the JB's operation. The Markdown and JNB files that will be used to build the JB HTML site can, and must, be placed within this directory in order to build the JB as well. This means that if changes are required to be made to the JB, one only needs to modify, or add too, the existing JNB and Markdown files present before rebuilding the JB. There are two main ways of viewing built JBs:

\begin{itemize}[leftmargin=.15in]
\item JBs can be viewed locally by opening the index.html file (generated after the JB has been built) found in the \_build/html folder of the JB directory in the user's browser of choice.
\item JBs can be published and viewed online by first placing the books source content on a public repository such as GitHub. GitHub Pages (GitHub's website builder that uses HTML, CSS, and JavaScript files present in a GitHub repository to publish websites) can then be used to build and make the source content available, to be viewed online (\textit{Publish Your Book Online}, n.d.).
\end{itemize}

\begin{table*}[ht]
    \begin{center}
        \begin{tabular}{ |p{6cm}|p{6cm}| }
        \hline
        \multicolumn{2}{|c|}{\textbf{Hosting Costs for a Repository Website}} \\
        \hline
        \textbf{Service} & \textbf{Cost}\\
        \hline
        Digital Ocean & \$174.00 CAD per year\\
        \hline
        Domain Name (NameCheap) & \$21.95 CAD per year \\
        \hline
        GitHub & \$0.00 CAD per year \\
        \hline
        \textbf{TOTAL COST} & \textbf{\$195.95 CAD per year} \\
        \hline
        \end{tabular}
        \caption{Hosting Costs for a Repository Website}
        \label{Zero 2 Table}
        \end{center}
\end{table*}

As JBs contain both the built HTML files and the source Markdown and JNB files, they are also capable of all the same functionalities and benefits of traditional JNBs. By directing Jupyter Labs, or one's preferred editor, to the JB directory on the local machine, the full functionality of the contained JNBs can be utilized. Any changes that the user wishes to make to the JB's Markdown and JNB source files can be reflected in the JB HTML site simply by saving the changes to the original Markdown and JNB files and rebuilding the JB. As well, unless embedded with large MP4 video files or images, JBs are generally quite small in total file size, especially when downloaded in a zip format. Instead of a collection of singular JNB and Markdown files that must each be opened individually in order to view, with JBs one can mimic the functionality of traditional textbooks, i.e. a singular viewable book containing all required information, while still allowing for the executability and modifiableness of JNBs. \\


\begin{center}
    {\Large \textbf{3.0 Jupyter-Materials-Repository Website}}
\end{center}

As has been outlined, the website needs to be designed and operated as an online repository of JMs, and this is the direction we are currently pursuing in our project with the University of Eswatini. A key aspect of a JM-repository website is in its simplicity; as the goal is to identify, target, and include low income areas of the world with limited access to the internet, a simple and lightweight website is ideal. Doing so would minimise costs such as loading time, bandwidth, and server costs while providing an avenue for students with little understanding of website design and maintenance to quickly learn how to modify and maintain websites in a low pressure and low stakes environment. The requirements of such a website are as follows:

\begin{figure*}[!ht]
    \begin{center}
    \includegraphics[scale=0.55]{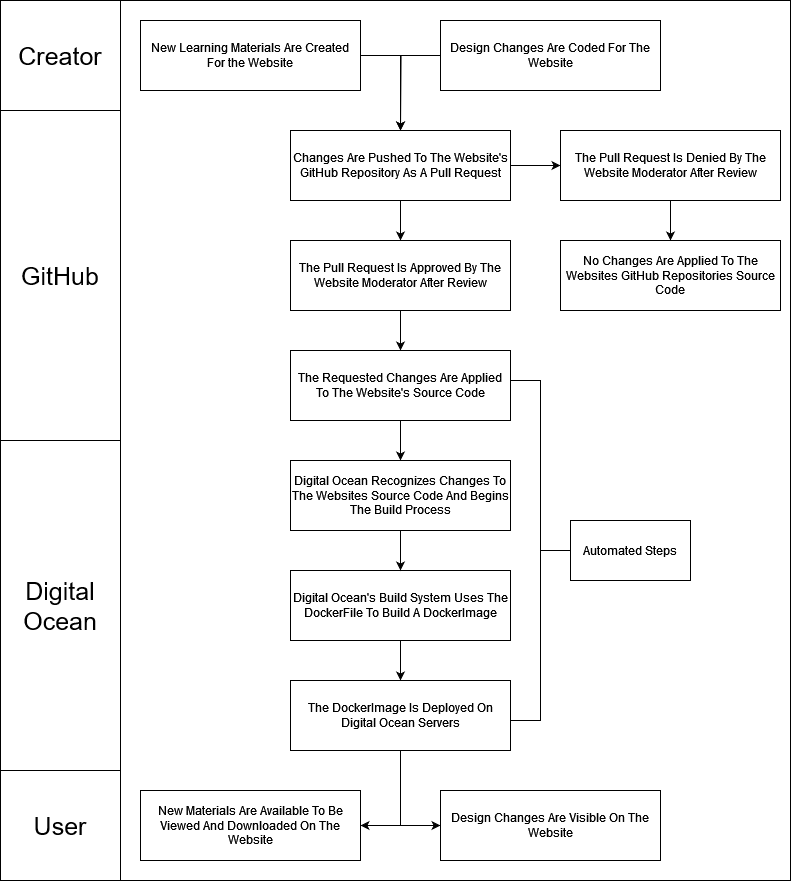}
    \captionof{figure}{User and Creator Experience Flow Chart}
    \end{center}
\end{figure*}

\begin{itemize}
\item The website must be built and hosted. Currently, there already exists a website in use for the University of Eswatini that can be forked\footnote{\noindent To fork a GitHub project is to copy the GitHub repositories contents into a new repository that can then be modified, edited, and added to without making any changes to the original repositories contents.} and modified for individuals needs; this website currently uses a combination of GitHub, DigitalOcean, and Docker\footnote{See Appendix A} to store, build, and host the website.
\item The website must be moderated so that material cannot be added at will to the website; unmoderated public repositories are vulnerable to unlawful or inappropriate materials being added.
\item Material must be created for the website. This requirement is two-fold: First, both instructors and students must be willing to ‘buy in’ to JM applications; this means a desire to create, and convert existing materials to JM-based textbooks and supplementary materials. Secondly, both instructors and students must be willing to learn how to work with Jupyter Labs and JNBs.
\item The website must provide (at least) the following functionality: (1) Store JMs and make them available to be downloaded. (2) Stored JMs must be made available to be viewed on the website. (3) The website must make available PDF and HTML versions of the JMs to be downloaded. (4) The website should be as lightweight as possible.
\end{itemize}

Currently, the JM-repository website developed for the University of Eswatini uses GitHub to store the website's source code, Docker to compile the websites source-code for the server, and Digital Ocean servers to to host the website. These applications work together as follows: 
\begin{enumerate}
  \item Changes, such as new JMs being added to the repository or design changes, are pushed\footnote{ To apply changes that have been made in one's local repository, to the GitHub repository} to GitHub using Git.
  \item Digital Ocean watches the GitHub repository and begins the automated build process when new changes to the repository are made.
  \item Digital Oceans build system uses the website's DockerFile to compile a DockerImage\footnote{See Appendix A} of the new state of the website.
  \item Digital Ocean deploys the newly compiled DockerImage on their servers at which point the new changes are visible on the website.
\end{enumerate}

\noindent
Note that other than pushing the desired changes to GitHub, and any moderation to ensure appropriateness, that the entirety of this process is automated once it has been established (see Figure 4). \\

GitHub's pull-request feature allows for new pushes to the GitHub repository, whether they be design changes or the addition of learning materials, to be reviewed before being pushed to the website's GitHub repository. In doing so, inappropriate/unsuitable learning materials or design changes that would effect the functionality of the website can be identified and rejected. As this feature is included natively as a functionality of GitHub, this provides a zero-cost maintenance and moderation option for the website. \\

As the goal for the website is to remain as simple and lightweight as possible in order to increase its viability in areas with poor access to the internet, this makes it ideal for students who have little to no experience in website design and maintenance to contribute to the continuation and addition of new functionalities to the website. The website itself can provide opportunities for learning and exploration for students, while simultaneously lowering the time and cost commitment of moderation and maintenance that may dissuade instructors if taken on alone. \\

Of course, there are costs associated with the hosting of any website as is shown in Table 1. Digital Ocean provides hosting at a cost of \$14.50 CAD per month, or \$174.00 CAD a year, while GitHub is free to use (\textit{Pricing Overview}, n.d.). Domain name prices will vary depending on which domain is chosen; '.com' domains can go for as little as \$21.95 CAD per year, with much cheaper but less recognizable options available as well (\textit{Domain Name Prices}, n.d.). \\


\begin{center}
    {\Large \textbf{4.0 Applications and Benefits}}
\end{center}

The applications of a JM-based repository website are as follows: 

\begin{itemize}[leftmargin=.15in]
\item Small scale repositories can center around materials for students in individual classrooms or university programs.
\item Large scale repositories can center around sharing materials from multiple universities for instructors to use.
\item The application of one, or both, of the above two points can service the needs of remote communities.
\end{itemize}

JM-based centralised repositories can facilitate the transfer of knowledge and expertise intra-disciplinarily as well as throughout universities, states/provinces, and countries; open-source class materials such as assignments, subject lessons, and textbooks can be shared between instructors teaching similar subjects. Currently, the process of procuring and using closed-source textbooks and materials has many issues:

\begin{itemize}[leftmargin=.15in]
\item Closed-source textbooks usually are copyrighted material; updates to existing information cannot legally be integrated into existing closed-source textbooks without express permission resulting in institutions being required to periodically purchase new textbooks to ensure correctness and relevance of subject matter.
\item The high cost of closed-source textbooks impacts student access to textbooks (with a majority not purchasing required textbooks), learning outcomes, and completion success of their studies (Florida Virtual Campus’s Office of Distance Learning and Student Service, 2016).
\item Even electronic versions of textbooks often include an annual payment model, or include some capability to restrict access after some period of time resulting in upfront costs being paid periodically and leading to higher costs over a period of time (Robinson et al., 2014). 
\end{itemize}

Open-source JM textbooks solve these issues; their upfront cost consists of the time it takes to create them or convert existing materials into the JNB format. Funds saved by school districts and universities by using open-source textbooks created in-house can be reinvested into their budgets, while providing instructors with the autonomy to modify their teaching materials to fit their particular needs instead of being forced into a specific path by the strict nature of closed-source textbooks. In this context, JM-based repository websites represent a unique opportunity for the development of open educational resources.\\

Furthermore, providing a platform for students to easily see and discuss the approaches their classmates employed to solve a given problem, exposes students to new learning practices in an easy to access manner and in a comfortable and safe environment. Students internalise the reasoning behind the solutions to problems in a greater capacity when exposed to different  methods for solving the problem at hand; having a wide array of approaches to a given problem allows students to have the opportunity to attempt different methods and develop a collection of tools when it comes to their reasoning (Mueller et al., 2010). \\

The exploration of alternative methods that can be used to solve problems, along with the internalisation of the knowledge required to understand such alternative methods, is facilitated by group discussion. Students who are able to "compare and challenge ideas and explanations" are better equipped to "recognize limitations, anomalies, and fallacies" about their internalisation of concepts and values. Environments which promote the concept of a "community of learners", wherein the students and teachers share the responsibility as it pertains to both learning and knowing, permit students to make meaningful connections to the course material (Mason, 1998). The use of JBs in break-out groups supports such group discussions during class time.\\


\begin{center}
    {\Large \textbf{5.0 Use in Areas with Limited Internet Access}}
\end{center}

One of the main appeals of a centralised JM website is its use in areas where students and instructors may be limited by poor internet access or speed, such as northern Canada and other less connected areas of the world. Climate challenges pertaining to the maintenance and operation of existing infrastructure, the distance of many remote communities from more developed areas, along with the distance between the communities themselves all limit the degree to which reliable internet access can be delivered (McMahon \& Akçayır, 2022). \\

Focusing on Canada as an example, the quality of education in its northern parts has been found to be lacking. A 2022 report (Task Force on Northern Post-Secondary Education, 2022) published by a Canadian Federal Task Force, investigated the northern post-secondary education sector and points to several issues hampering these northern communities, namely: the “...reading, writing, maths skills in our students are generally low”, and the “ongoing digital divide” appearing in Canada is playing a large role in this decline in skills needed for successfully participating in post-secondary education. As stated in the report “...learning management systems, asynchronous or synchronous options, and sometimes even email, cannot be relied upon consistently.” Beyond that, cultural differences in lifestyle and accommodations play a larger role in the success of students from remote northern communities adjusting to schools than those from less remote areas of Canada (Task Force on Northern Post-Secondary Education, 2022). These issues are not unique to Canada; disparity in learning outcomes can be found in remote communities around the world (Gardiner, 2008; Guenther et al., 2014; Nworgu \& Nworgu, 2013). \\

A lightweight repository website helps mitigate issues surrounding the accessibility to new and powerful tools, such as JMs, in areas of the world with poor internet access. An important aspect of JMs is that they only need to be downloaded once. If a student wishes to make changes, they can make a copy of the JM once to ensure they have a copy of the original. This means that once they have Jupyter Labs installed, along with any packages they know they may require, students can download any needed JMs contained in the repository website from a centralised location and continue to study and work with the full functionality of both Jupyter Labs and JMs while being offline, no matter the location, including the running of code. With Jupyter Lab's offline capabilities, and its open-source and cost-free nature, JNBs provide a tremendous amount of value to low-income students without access to a stable internet connection. \\


\begin{center}
    {\Large \textbf{6.0 Conversion of Materials to Jupyter Notebooks and PDF/HTML}}
\end{center}

While there are hurdles associated with setting up the repository website such as cost, maintenance, and moderation, the main hurdle for a project such as this is getting the individuals who would use and benefit from it to buy into the project. Getting instructors to start providing content for the website will revolve around the conversion of their existing materials, which will have served them well and are in a format they are comfortable with, to a new format they most likely have not used before. While, ideally, instructors would be adequately trained and proficient in how to use JMs and how to format text with Markdown, tools exist that can be used to mitigate any lack of knowledge instructors may have concerning conversion. These tools are largely free to use, although full functionality is often locked behind a paywall. \\

Writage, a markdown converter plugin for Microsoft Word, makes it easy to convert existing Word documents into Markdown if one is not comfortable with Markdown formatting (\textit{Writage}, n.d.). Using Writage is as simple as downloading the plugin, then saving any existing documents using the Markdown file type instead of .doc/.docx. The converted document can then be opened using notepad, or some other text editor, and copy \& pasted into a JNB Markdown cell. Writage has some issues when it comes to converting any embedded images in the original document, but JNBs allows for easy and simple methods of adding images, such as drag-and-dropping images right into the cell. \\

\begin{figure*}[ht]
    \begin{center}
    \includegraphics[scale = 0.7]{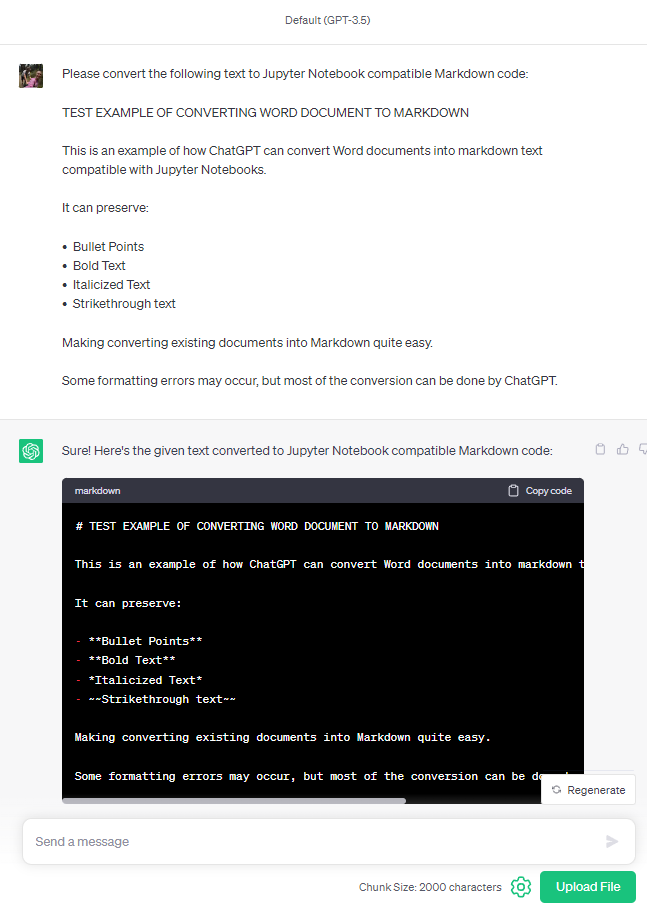}
    \captionof{figure}{Converting Word Document to Markdown using ChatGPT}
    \end{center}
\end{figure*}

OpenAI’s ChatGPT offers another method for converting formatted text into Markdown (\textit{OpenAI Platform}, n.d.). Instructors can copy \& paste text into the chat-bot and have ChatGPT convert it to Markdown, with little Markdown knowledge required. There also exists a chrome plugin, ChatGPT File Uploader, that allows the user to select a file on their computer to be uploaded; the plugin feeds the text from the document into ChatGPT and converts it to Markdown with the appropriate prompt (\textit{ChatGPT File Uploader}, n.d.). Like Writage, ChatGPT will most likely not be able to correctly embed images into the Markdown text. However, as previously mentioned, adding images to JNBs using the drag-and-drop method is quite simple. \\

\begin{table*}[ht]
    \begin{center}
        \begin{tabular}{ |p{3cm}|p{1cm}|p{3cm}|p{1cm}|p{2cm}|p{2cm}|p{2cm}|  }
        \hline
        \multicolumn{7}{|c|}{\textbf{Raspberry Pi Model Specifications}} \\
        \hline
        \textbf{Model} & \textbf{Ram} & \textbf{Processor} & \textbf{WIFI} & \textbf{Bluetooth} & \textbf{USB Ports} & \textbf{Cost}\\
        \hline
        Raspberry Pi 4 Model B/1GB & 1GB &  Quad core Cortex-A72 (ARM v8) 64-bit SoC @ 1.5GHz & Yes & Yes & 4 & \$48.95 CAD \\
        \hline
        Raspberry Pi 4 Model B/2GB & 2GB &  Quad core Cortex-A72 (ARM v8) 64-bit SoC @ 1.5GHz & Yes & Yes & 4 & \$62.95 CAD \\
        \hline
        Raspberry Pi 4 Model B/4GB & 4GB &  Quad core Cortex-A72 (ARM v8) 64-bit SoC @ 1.5GHz & Yes & Yes & 4 & \$76.95 CAD \\
        \hline
        Raspberry Pi 400 & 4GB &  Quad core Cortex-A72 (ARM v8) 64-bit SoC @ 1.8GHz & Yes & Yes & 3 & \$98.95 CAD \\
        \hline
        Raspberry Pi 4 Model B/8GB & 8GB &  Quad core Cortex-A72 (ARM v8) 64-bit SoC @ 1.5GHz & Yes & Yes & 4 & \$104.95 CAD \\
        \hline
        \end{tabular}
        \caption{Raspberry Pi Model Specifications}
        \label{Raspberry Pi Model Specifications}
    \end{center}
\end{table*}

\begin{table*}[ht]
    \begin{center}
        \begin{tabular}{ |p{5.5cm}|p{4cm}|p{4cm}|  }
            \hline
            \multicolumn{3}{|c|}{\textbf{Raspberry Pi Peripherals Cost}} \\
            \hline
            \textbf{Peripheral} & \textbf{Peripheral Cost} & \textbf{Peripheral Source}\\
            \hline
            Bluetooth Mouse & \textasciitilde{\$10.00 CAD} & staples.ca \\
            \hline
            Bluetooth Keyboard & \textasciitilde{\$20.00 CAD} & staples.ca \\
            \hline
            USB-C Power Supply & \$9.95 CAD & pishop.ca \\
            \hline
            micro-SD Card RB Pi OS & \$14.95 CAD & pishop.ca \\
            \hline
            micro-HDMI to HDMI-A cord & \$5.95 CAD & pishop.ca \\
            \hline
            \textbf{TOTAL COST} & \multicolumn{2}{|l|}{\textbf{\textasciitilde{\$60.85 CAD}}} \\
            \hline
        \end{tabular}
        \caption{Raspberry Pi Peripherals Cost}
        \label{Raspberry Pi Peripherals Cost}
    \end{center}
\end{table*}

With these tools, conversion of existing materials can be done easily and quickly, lowering the cost of entry to making the switch from traditional Word and PDF documents to JNBs. Along with these solutions, Jupyter also provides tools for converting JNBs to PDF and HTML in the form of nbconvert. Currently, nbconvert supports the conversion of JNBs to HTML, LaTex, PDF, WebPDF, Reveal.js, HTML Slideshow, Markdown, ASCII, reStructuredText, executable script and notebook (\textit{Using as a Command Line Tool}, n.d.). However, nbconvert only supports conversion to HTML natively; if an instructor or student wishes to convert their JNB to formats other than HTML, they will have to also install Pandoc, Tex, and Pyppeteer as separate installations (\textit{Installation — Nbconvert}, n.d.). These conversion methods give instructors and students a tremendous amount of control over the final product; code cells can be cut off or line-wrapped, font style and size can be varied, output from code cells can be included or removed, etc. The choice to switch to using JNBs does not preclude the use of PDF documents as a method of distribution. In fact, using JNBs allows for instructors and students to enhance their PDFs with stylish formatting using Markdown and the option to include the output from any code present in the document. \\


\begin{center}
    {\Large \textbf{7.0 Integration of Raspberry Pi Kits}}
\end{center}

\begin{table*}[ht]
    \begin{center}
        \begin{tabular}{ |p{5.5cm}|p{4cm}| }
        \hline
        \multicolumn{2}{|c|}{\textbf{Raspberry Pi Model Cost Comparison w/ Peripherals}} \\
        \hline
        \textbf{Item} & \textbf{Cost}\\
        \hline
        Raspberry Pi 4 Model B/1GB & \$48.95 CAD \\
        \hline
        Peripherals & \$60.85 CAD \\
        \hline
        \textbf{TOTAL COST} & \textbf{\$109.80} \\
        \hline
        \rowcolor{gray!50}
        \multicolumn{2}{|c|}{} \\
        \hline
        Raspberry Pi 4 Model B/2GB & \$62.95 CAD \\
        \hline
        Peripherals & \$60.85 CAD \\
        \hline
        \textbf{TOTAL COST} & \textbf{\$123.80} \\
        \hline
        \rowcolor{gray!50}
        \multicolumn{2}{|c|}{} \\
        \hline
        Raspberry Pi 4 Model B/4GB & \$76.95 CAD \\
        \hline
        Peripherals & \$60.85 CAD \\
        \hline
        \textbf{TOTAL COST} & \textbf{\$137.80} \\
        \hline
        \rowcolor{gray!50}
        \multicolumn{2}{|c|}{} \\
        \hline
        Raspberry Pi 4 Model B/8GB & \$104.95 CAD \\
        \hline
        Peripherals & \$60.85 CAD \\
        \hline
        \textbf{TOTAL COST} & \textbf{\$165.80} \\
        \hline
        \rowcolor{gray!50}
        \multicolumn{2}{|c|}{} \\
        \hline
        Raspberry Pi 400 - Complete Kit & \$136.95 CAD \\
        \hline
        Peripherals & Included \\
        \hline
        \textbf{TOTAL COST} & \textbf{\$136.95} \\
        \hline
        \end{tabular}
        \caption{Raspberry Pi Model Cost Comparison w/ Peripherals}
        \label{Raspberry Pi Model Cost Comparison w/ Peripherals}
    \end{center}
\end{table*}

\begin{table*}[hb]
    \begin{center}
        \begin{tabular}{ |p{8cm}|p{4cm}| }
        \hline
        \multicolumn{2}{|c|}{\textbf{Raspberry Pi Full Package Price}} \\
        \hline
        \textbf{Raspberry Pi Model} & \textbf{Package Cost}\\
        \hline
        Raspberry Pi 4 Model B/1GB Kit w/ Monitor & \$184.80 CAD \\
        \hline
        Raspberry Pi 4 Model B/2GB Kit w/ Monitor & \$198.80 CAD \\
        \hline
        Raspberry Pi 4 Model B/4GB Kit w/ Monitor & \$212.80 CAD \\
        \hline
        Raspberry Pi 4 Model B/8GB Kit w/ Monitor & \$240.80 CAD \\
        \hline
        Raspberry Pi 400 Package w/ Monitor & \$211.95 CAD \\
        \hline
        \end{tabular}
        \caption{Raspberry Pi Full Package Price}
        \label{Raspberry Pi Full Package Price}
    \end{center}
\end{table*}

While unreliable internet access and speed can still be an impediment for a web repository project, access to affordable personal computers is also a limiting factor for both delivering an acceptable level of education and providing an outlet for personal intellectual exploration. Students who have access to a home computer are found to have a 6-8 percentage point increase in the probability of graduating over those who do not (Fairlie et al., 2010). Furthermore, the cost of living in the northern parts of Canada, for example, has been found to be 1.46 times that of the southern areas of Canada, and the incidence of poverty in the Canadian north for families with children is 31.3\% compared to 9.9\% in the South (Daley et al., 2015). \\

This cost of living discrepancy is compounded and exasperated by the lack of cheap and large scale shipping methods available to Northern communities. Despite accounting for almost 40\% of Canada’s landmass, only approximately 1\% of Canada’s road network and 0.3\% of Canada’s rail lines are located in the three (northern) territories. The lack of road and rail transportation leads to many cities relying on goods being brought in by aeroplane, a much more expensive alternative (\textit{Transportation in the North}, 2008). As such, there exists a demand for inexpensive products that can perform as adequately as their more expensive counterparts in order to offset the additional costs incurred when shipping to areas such as Northern Canada. These same issues exist in many parts of the world; shipping to, and within, smaller and more remote countries also incurs major additional costs. \\

Raspberry Pis\footnote{See Appendix A} are currently well suited to mitigate these issues when it comes to personal computers. Their benefits include:

\begin{itemize}[leftmargin=.15in]
\item Small and light, weighing \textasciitilde{150 grams} and measuring roughly 9 x 6 x 2 cm, resulting in competitive bulk shipping costs, especially compared to more traditional computers.
\item Raspberry Pi kits are themselves inexpensive compared to traditional computers (see Table 2). This includes both the Raspberry Pi itself as well as any supplementary or replacement hardware.
\item Outside of computationally heavy processes and projects (playing video games, running large programs such as AutoCAD, etc), Raspberry Pis are largely able to perform the same tasks as traditional computers.
\end{itemize}

\begin{table*}[ht]
    \begin{center}
        \begin{tabular}{ |p{5.5cm}|p{2cm}|p{2cm}|p{2cm}|}
        \hline
        \multicolumn{4}{|c|}{\textbf{Raspberry Pi Price Comparison with ASUS Chromebook C204MA Q14}} \\
        \hline
        \textbf{Raspberry Pi Model} & \textbf{Raspberry Pi Cost} & \textbf{Chromebook Cost} & \textbf{Price Difference (\%)}\\
        \hline
        Raspberry Pi 4 Model B/1GB Kit w/ Monitor & \$184.80 CAD & \$374.99 & 50.71\% \\
        \hline
        Raspberry Pi 4 Model B/2GB Kit w/ Monitor & \$198.80 CAD & \$374.99 & 46.99\% \\
        \hline
        Raspberry Pi 4 Model B/4GB Kit w/ Monitor & \$212.80 CAD & \$374.99 & 43.25\% \\
        \hline
        Raspberry Pi 4 Model B/8GB Kit w/ Monitor & \$240.80 CAD & \$374.99 & 35.78\% \\
        \hline
        Raspberry Pi Package w/ Monitor & \$211.95 CAD & \$374.99 & 43.48\% \\
        \hline
        \end{tabular}
        \caption{Raspberry Pi Price Comparison with ASUS Chromebook C204MA Q14}
        \label{Raspberry Pi Price Comparison with ASUS Chromebook C204MA Q14}
    \end{center}
\end{table*}

\begin{table*}[hb]
    \begin{center}
        \begin{tabular}{ |p{2cm}|p{4cm}|p{5cm}|}
        \hline
        \multicolumn{3}{|c|}{\textbf{Raspberry Pi Spec Comparison with ASUS Chromebook C204MA Q14}} \\
        \hline
        & \textbf{Raspberry Pi 4 Model B} & \textbf{ASUS Chromebook C204MA}\\
        \hline
        Processor & Quad-Core 1.5GHz & Dual-Core 1.1GHz \\
        \hline
        RAM & 1/2/4/8GB & 4GB \\
        \hline
        Memory & Up to 1TB; dependant on micro-SD card size & 32GB \\
        \hline
        USB Ports & 4 & 2 \\
        \hline
        USB-C Ports & 0 & 2 \\
        \hline
        Bluetooth & Yes & Yes \\
        \hline
        WIFI & Yes & Yes \\
        \hline
        \end{tabular}
        \captionof{table}{Raspberry Pi Spec Comparison with ASUS Chromebook C204MA Q14}
        \label{Raspberry Pi Spec Comparison with ASUS Chromebook C204MA Q14}
    \end{center}
\end{table*}

As mentioned, one of the main benefits of Raspberry Pis are their low price point. This stems from stripping away 'non-essential' components compared to traditional computers. The Raspberry Pi 4 Model B is currently the newest model available and ranges from \$48.95 to \$104.95 (\textit{Current Raspberry Pi Boards in Canada}, n.d.). \\

Additionally, Raspberry Pis, as with traditional computers, require peripherals in order to function. Items such as a mouse, keyboard, power supply and HDMI cord would need to be shipped alongside the Raspberry Pi. As can be seen in Table 3, the cost of purchasing all the required peripherals for the Raspberry Pi adds approximately \textasciitilde{\$60.85} CAD. However, as mentioned, Raspberry Pi also sells a package for the Raspberry Pi 400 which includes all the peripherals. The cost of a sourced Raspberry Pi kit, as seen in Table 4, is then as low as \$109.80 CAD for the Model B 1GB model, up to \$165.80 CAD for the Model B 8GB Model, and \$136.95 for the Raspberry Pi 400 (Plo, n.d.). \\

Both the prepackages and peripheral kits, however, would also require the purchase of a monitor, or for the user to already own a monitor/TV to connect the Raspberry Pi to. Refurbished monitors can be found online often in the range of \textasciitilde{\$50 - \textasciitilde{\$100 CAD}} (\textit{Cheap Used and Refurbished Monitors}, n.d.). Separating the monitor and the computer also allows for hardware pieces to be replaced easily compared to traditional laptops. Raspberry Pis' low price point also makes buying and storing extra units in case of hardware or software issues a much more feasible task. Taking an averaged price for refurbished monitors, \textasciitilde{\$75 CAD}, the cost for creating a Raspberry Pi kit can be seen in Table 5. \\

Note that these prices assume that everything is bought at retail prices; wholesale deals with manufacturers, an option for universities, colleges and school boards, could bring down the price of monitors and peripherals considerably. Even so, when compared with lower-end Chromebook models, which are often purchased in large quantities by school districts to facilitate online learning (“Chromebooks En Route,” 2021), Raspberry Pis compare quite well in both specifications and price. This can be shown using the ASUS Chromebook C204MA Q14 as a comparison model (see Table 6) (\textit{ASUS ChromeBook}, n.d.). \\

As it is demonstrated in the specification comparison in Table 7, Raspberry Pis match or exceed the ASUS Chromebook C204MA Q14 in most categories. Furthermore, as shown in Table 6, we can see a price difference of up to \textasciitilde{50\%}; twice as many Raspberry Pi kits could be purchased and distributed as compared to the ASUS Chromebook C204MA Q14 for the same expenditure. When considering mass purchasing units in order to supply an entire university, college, or school, the cost-savings become quite apparent. The only major downside to Raspberry Pis compared to laptops is in its portability; traveling with a Raspberry Pi also means traveling with all the peripherals and the monitor required for its operation, or having them available at point-of-use. \\

Moreover, there is no difference in installing Jupyter Labs on Raspberry Pi units compared to traditional computers. While the Raspberry Pi does run a Linux based operating system, Jupyter Labs is easily installed using Pythons package installer, Pip; the commands and process for installing Jupyter Labs are the same on a Raspberry Pi as they would be on a traditional computer. Raspberry Pi 4 Model B 2GB edition does an excellent job of running Jupyter Labs, along with large scale JNBs. As would be expected, the Raspberry Pi 400, which comes with 4GB of RAM, also performs excellent, having no issues with running both Jupyter Labs and JNBs. All-in-all, JNBs and Raspberry Pis work together quite well, making Raspberry Pis an excellent tool to be used in conjunction with a JNB repository website. \\


\begin{center}
    {\Large \textbf{8.0 Conclusion and Outlook}}
\end{center}

JMs have been shown to facilitate and improve learning outcomes in a variety of computational subjects for a wide range of students. Notably, JB's ability to provide a traditional textbook experience, combined with interactivity and functionality built into the learning material, and an environment that can be operated offline as well as online, allows for a much more direct hands-on learning experience. These benefits, in conjunction with their ability to be converted into conventional formats (e.g. PDFs) deliver a tremendous amount of utility and freedom to instructors as to how they choose to deliver classroom materials. Furthermore, their open-source nature offers the opportunity for collaboration between instructors regardless of their institution or location. \\

Additionally, a repository website centered around the collection and dissemination of JM learning materials can work to serve both individual classrooms as well as larger groups of institutions. The collaboration and the sharing of materials can be facilitated by learning materials created in one institution but utilized or improved upon elsewhere by other instructors or students. The benefits of the collaboration facilitated by a repository website can be used to improve the learning outcomes in remote communities where the availability of subject-matter expertise can be limited. Moreover, website maintenance and creation is becoming easier and easier, whether one considers building a website from scratch as is outlined in this paper, or using website builder services such as WordPress. \\

Raspberry Pis can also provide a tremendous amount of value when used with Jupyter materials. Due to their modest cost, reasonable performance, and low cost of shipping when compared to traditional desktop and laptop computers, Raspberry Pis are uniquely situated to provide computing power and accessibility for low-income and remote communities. Their ability to operate JMs also means that when paired with a JM repository website, learning outcomes can, in some measure, be normalized regardless of location or income, providing benefits to both students and instructors. \\

The cheap processing power of the Raspberry Pi, combined with the offline functionality and open-source nature of JMs, means that regardless of physical location, income, or quality of internet accessibility, students and instructors, through a JM repository, can work with the same learning materials. Additionally, students, irrespective of location, will benefit from the utility and functionality of JMs as compared to traditional learning materials. This levelling of the playing field of both the accessibility and quality of learning materials has the opportunity to lead to increased learning outcomes for students and, as such, should be strongly considered by universities and institutions around the world.

\end{multicols}


\newpage
\noindent
{\Large \textbf{Bibliography}} \\


\begin{hangparas}{.25in}{1}

\textit{ASUS ChromeBook C204MA Q1R - 11.6" - Intel Celeron - N4020 - 4 GB RAM - 32 GB EMMC - C204MA-Q1R-CB - Laptops - CDW.CA}. (n.d.). CDW.CA. https://www.cdw.ca/product/asus-chromebook-c204ma-q1r-11.6-intel-celeron-n4020-4-gb-ram-32/6390988?pfm=srh \\

\textit{Binder Documentation — Binder 0.1b documentation}. (n.d.). https://mybinder.readthedocs.io/en/latest/ \\

\textit{Build your book}. (n.d.). https://jupyterbook.org/en/stable/start/build.html \\

Castilla, R., \& Peña, M. (2023). Jupyter Notebooks for the study of advanced topics in Fluid Mechanics. \textit{Computer Applications in Engineering Education, 31}(4), 1001–1013. doi:10.1002/cae.22619 \\

\textit{ChatGPT File Uploader}. (n.d.). https://chromewebstore.google.com/detail/chatgpt-file-uploader/oaogphgfdbdbmhkiplemgehihiiececj?pli=1 \\

\textit{Cheap used and refurbished monitors | Discount Computer Depot}. (n.d.). https://discountcomputerdepot.com/categories/refurbished-monitors.html \\

Chromebooks en route as N.L. schools move online, says education minister. (2021, February 15). \textit{CBC}. https://www.cbc.ca/news/canada/newfoundland-labrador/education-schools-daycare-covid-update-1.5914330 \\

\textit{Current Raspberry Pi boards in Canada}. (n.d.). https://www.pishop.ca/product-category/raspberry-pi/raspberry-pi-boards/current-pi-boards/ \\

Daley, A., Burton, P., \& Phipps, S. (2015). Measuring poverty and inequality in northern Canada. \textit{Journal of Children and Poverty, 21}(2), 89–110. doi:10.1080/10796126.2015.1089147 \\

Daring Fireball: Markdown. (n.d.). https://daringfireball.net/projects/markdown/ \\

\textit{Domain name prices | Domain registration costs — Namecheap}. (n.d.). https://www.namecheap.com/domains/\#pricing \\

Fairlie, R. W., Beltran, D. O., \& Das, K. K. (2010). Home computers and educational outcomes: Evidence from the NLSY97 and CPS. \textit{Economic inquiry, 48}(3), 771-792. \\

Fischer, L., Hilton, J., Robinson, T. J., \& Wiley, D. (2015). A multi-institutional study of the impact of open textbook adoption on the learning outcomes of post-secondary students. \textit{Journal of Computing in Higher Education}, 27(3), 159–172. doi:10.1007/s12528-015-9101-x \\

Florida Virtual Campus’s Office of Distance Learning and Student Service. (2016). \textit{2016 Student Textbook and Course Materials Survey.} https://www.oerknowledgecloud.org/archive/2016\%20Student\%20Textbook\%20Survey.pdf \\

Gardiner, M. (2008). Education in rural areas. \textit{Issues in education policy, 4}, 1-33. \\

\textit{Google Colaboratory}. (n.d.). https://colab.research.google.com/github/coolernato/Introduction-to-Python/blob/master/Using\%20Jupyter\%20Notebooks.ipynb \\

Guenther, J., Bat, M., \& Osborne, S. (2014). Red dirt thinking on remote educational advantage. Australian and International Journal of Rural Education, 24(1), 51–67. doi:10.3316/informit.197109386575105 \\

\textit{Installation — nbconvert 7.14.0 documentation}. (n.d.). https://nbconvert.readthedocs.io/en/latest/install.html \\

Isihara, P. (2021). A College Teacher’s Introduction to Jupyter Notebooks. \textit{International Journal for Technology in Mathematics Education, 28}(4), 235–244. doi:10.1564/tme\_v28.4.03 \\

Jupyter Notebooks — Anaconda documentation. (n.d.). https://server-docs.anaconda.com/en/latest/user/notebook.html \\

\textit{JupyterLab Documentation — JupyterLab 4.0.9 documentation}. (n.d.). https://jupyterlab.readthedocs.io/en/stable/ \\

\textit{Kernels (Programming Languages) — Jupyter Documentation 4.1.1 alpha documentation}. (n.d.). https://docs.jupyter.org/en/latest/projects/kernels.html \\

Kurniawan, A. (2018). \textit{Raspbian OS Programming with the Raspberry Pi: IoT Projects with Wolfram, Mathematica, and Scratch}. Apress. \\

\textit{Markdown Cells — Jupyter Notebook 7.0.6 documentation}. (n.d.). https://jupyter-notebook.readthedocs.io/en/stable/examples/Notebook/Working\%20With\%20Markdown\%20Cells.html \\

Mason, L. (1998). Sharing cognition to construct scientific knowledge in school context: The role of oral and written discourse. \textit{Instructional Science, 26}(5), 359–389. \\

McMahon, R., \& Akçayır, M. (2022). Investigating concentrated exclusion in telecommunications development: Engaging rural voices from Northern Canada. Journal of Rural Studies, 95, 183–194. doi:10.1016/j.jrurstud.2022.09.004 \\

Mitsioni, M. F., Stouras, M., \& Makedonas, C. (2023). Taking School Instrumentation One Step Forward: A Do-It-Yourself Type Spectrophotometer and a Jupyter Notebook That Enable Real Time Spectroscopy during School Lessons. \textit{Journal of Chemical Education, 100}(7), 2704–2712. doi:10.1021/acs.jchemed.3c00248 \\

Mueller, M., Yankelewitz, D., \& Maher, C. A. (2010). Rules without Reason: allowing students to rethink previous conceptions. \textit{The Mathematics Enthusiast, 7}(2–3), 307–320. doi:10.54870/1551-3440.1190 \\

Mushtaha, E., Abu Dabous, S., Alsyouf, I., Ahmed, A., \& Raafat Abdraboh, N. (2022). The challenges and opportunities of online learning and teaching at engineering and theoretical colleges during the pandemic. \textit{Ain Shams Engineering Journal}, 13(6). doi:10.1016/j.asej.2022.101770 \\

\textit{nbconvert: Convert Notebooks to other formats — nbconvert 7.14.0 documentation}. (n.d.). https://nbconvert.readthedocs.io/en/latest/index.html \\

\textit{nbgrader — nbgrader 0.9.1 documentation}. (n.d.). https://nbgrader.readthedocs.io/en/stable/index.html \\

Noprianto, Wijayaningrum, V. N., \& Lestari, V. A. (2022). Jupyter Lab Platform-Based Interactive Learning. \textit{IEEE}. doi:10.1109/IEIT56384.2022.9967857 \\

Nworgu, B. G., \& Nworgu, L. N. (2013). Urban–rural disparities in achievement at the basic education level: the plight of the rural child in a developing country. \textit{Journal of Developing Country Studies. 3}(14). p. 128, 139. \\

\textit{Open Educational Resources (OER) in Canada | Centre for Teaching Excellence}. (n.d.). https://uwaterloo.ca/centre-for-teaching-excellence/oer-canada \\

Open UToronto. (2023, October 18). \textit{Open Resources - Open UToronto}. Open UToronto - Open Resources and Innovation Projects at the University of Toronto. https://ocw.utoronto.ca/open-resources/ \\

\textit{OpenAI Platform}. (n.d.). https://platform.openai.com/docs/introduction \\

Perkel, J. M. (2018). Why Jupyter is data scientists’ computational notebook of choice. \textit{Nature, 563}(7729), 145–146. doi:10.1038/d41586-018-07196-1 \\

Pillay, C. S. (2020). Analyzing biological models and data sets using Jupyter notebooks as an alternate to laboratory‐based exercises during COVID‐19. \textit{Biochemistry and Molecular Biology Education, 48}(5), 532–534. doi:10.1002/bmb.21443 \\

Plo. (n.d.). \textit{Raspberry Pi 400 - Complete Kit}. PiShop.ca. https://www.pishop.ca/product/raspberry-pi-400-complete-kit/ \\

\textit{Pricing Overview | DigitalOcean}. (n.d.). https://www.digitalocean.com/pricing \\

\textit{Project Jupyter}. (n.d.-a). Home. https://jupyter.org/ \\

\textit{Project Jupyter}. (n.d.-b). About Us. https://jupyter.org/about \\

\textit{Project Jupyter Documentation — Jupyter Documentation 4.1.1 alpha documentation}. (n.d.). https://docs.jupyter.org/en/latest/ \\

\textit{Publish your book online}. (n.d.). https://jupyterbook.org/en/stable/start/publish.html

Quinn, P. (2016). \textit{Google schools? A chromebook case study. Screen Education, 82} \\

Raspberry Pi documentation. (n.d.). https://www.raspberrypi.com/documentation/ \\

Reades, J. (2020). Teaching on Jupyter. \textit{Region, 7}(1), 21–34. doi:10.18335/region.v7i1.282 \\

Robinson, T. J., Fischer, L., Wiley, D., \& Hilton, J. (2014). The Impact of Open Textbooks on Secondary Science Learning Outcomes. \textit{Educational Researcher, 43}(7), 341–351. \\

Ruiz-Sarmiento, J., Baltanas, S., \& González-Jiménez, J. (2021). Jupyter Notebooks in Undergraduate Mobile Robotics Courses: Educational tool and case study. \textit{Applied Sciences, 11}(3), 917. doi:10.3390/app11030917 \\

Task Force on Northern Post-Secondary Education. (2022). \textit{A Shared Responsibility: Northern Voices, Northern Solutions—Report of the Task Force on Northern Post-Secondary Education.} https://northernpse.ca/sites/default/files/2022-05/16513\%20CIRNAC\%20Northern\%20PS\%20Edu\_AR\_EN\_2022May20.pdf \\

\textit{The Executable Books project}. (n.d.). Executable Book Project. https://executablebooks.org/en/latest/index.html \\

\textit{Transportation in the north}. (2008, March 26). https://www150.statcan.gc.ca/n1/pub/16-002-x/2009001/article/10820-eng.htm \\

\textit{Tutorials}. (n.d.). https://gwosc.org/tutorials/ \\

\textit{Using as a command line tool — nbconvert 7.14.0 documentation}. (n.d.). https://nbconvert.readthedocs.io/en/latest/usage.html \\

Weiss, C. J. (2020). A Creative Commons Textbook for Teaching Scientific Computing to Chemistry Students with Python and Jupyter Notebooks. Journal of Chemical Education, 98(2), 489–494. doi:10.1021/acs.jchemed.0c01071 \\

\textit{Why we use Jupyter notebooks | Teaching and Learning with Jupyter}. (2019, December 6). https://jupyter4edu.github.io/jupyter-edu-book/why-we-use-jupyter-notebooks.html \\

\textit{Working with Jupyter Notebooks in Visual Studio Code}. (2021, November 3). https://code.visualstudio.com/docs/datascience/jupyter-notebooks \\

\textit{Writage}. (n.d.). Writage - Markdown plugin for Microsoft Word | Documentation. https://www.writage.com/docs-page/\#section-1 \\

\end{hangparas}


\noindent
{\Large \textbf{Appendix A.}} \\

\noindent
This appendix consists of descriptions and explanations of the programs, applications, and website services used to build, host, and maintain the JM repository website for the University of Eswatini and the proposed website as outlined in this paper. \\

\noindent
{\large \textbf{Raspberry Pi}} \\

\noindent
The Raspberry Pi is an inexpensive, wallet sized computer that operates in a manner similar to a traditional computer. It requires a monitor, mouse, and keyboard for most uses, although it can be utilized as a motherboard or Single Board Computer (SBC) for applications such as Plex servers, video doorbells, robotics, etc. A simple way of understanding the Raspberry Pi is that it is a traditional computer that has been stripped down to its essentials, to what is only necessary, in order to be as light weight and inexpensive as possible. Raspberry Pi uses the Raspberry Pi OS (formerly known as Raspian) as its operating system for nearly every product available. \\

\noindent
{\large \textbf{GitHub}} \\

\noindent
GitHub is a website service that provides software and website developers online code repositories for code, projects, and instructions in a centralised location where multiple developers can access and modify the project. This makes large scale collaborative projects easy to manage and modify as well as allowing contributors to be located anywhere, provided they have internet access. GitHub makes use of Git, which is a distributed version control system (i.e.: the entirety of the codebase is available on each repository member's computer) that allows users to create duplicates of the source code, known as branching. This duplicated source code can be modified and tested before being merged into the main branch source code. This helps identify and mitigate issues before changes are applied to the project. Changes to the repository can be done on a line by line basis, meaning that several sections can be modified at a time without affecting other sections. As well, changes to the repository can be undone by reverting back to an earlier version of the project.\\

\noindent
{\large \textbf{Digital Ocean}} \\

\noindent
Digital Ocean is a cloud computing distributor that provides an 'infrastructure as a service' platform for software and website developers. Digital Ocean uses Kubernetes, a service that allows for multiple containers to operate simultaneously on one server, enabling the hosting of almost three million websites. Digital Ocean is able to be directed at code repositories, such as GitHub, and build and deploy the contained code on their servers, along with rebuilding and redeploying when changes have been made to the repository; once this process has been initialized, Digital Ocean is able to perform these tasks automatically. For the website in discussion in this paper, Digital Ocean watches the website's GitHub repository for any changes, using the contained DockerFile to build a DockerImage that is deployed on their servers in order to operate the website. \\

\noindent
{\large \textbf{Docker}} \\

\noindent
Docker is an open-source project that allows for the creation of self-contained containers used for the deployment of applications both in the cloud and on local devices. These containers allow for applications to run in isolated environments using packages and code unique to the application (i.e.: running a Linux container deploying a Linux based application on a Windows computer). As it relates to this project, the website's deployment requires a DockerImage, a read only file containing deployment instructions for creating a container, which is built using a DockerFile, a text file containing instructions for what to include in the DockerImage. Docker creates an image for each separate line in the DockerFile which are then compiled to create the DockerImage. Doing so allows for a quicker compilation when changes are made to the DockerFile, as only the lines below the modified line need to be recompiled. Once the DockerImage is compiled, it is used to create a container that deploys the code compiled in the DockerImage, in this case a container run on Digital Ocean servers containing the code for a website. 

\newpage

\noindent
{\large \textbf{List of Abbreviations}} \\

\noindent
Jupyter Book - JB \\

\noindent
Jupyter Notebook - JNB \\

\noindent
Jupyter Materials - JM \\

\newpage

\noindent
{\Large \textbf{Declarations}} \\

\noindent
{\large \textbf{Availability of Data and Materials}} \\

\noindent
Not applicable. \\


\noindent
{\large \textbf{Competing Interests}} \\

\noindent
The authors have no conflicts of interest to declare. \\


\noindent
{\large \textbf{Funding}} \\

\noindent 
LEFT BLANK \\


\noindent
{\large \textbf{Authors' Contributions}} \\

\noindent
Faculty of Mathematics and Science, Brock University, St. Catharines, ON, Canada

\noindent
Peter Berg  \\

\noindent 
Augustana Campus, University of Alberta, Camrose, AB, Canada

\noindent
Zachary Kelly \\

\noindent
Each author read and approved the final manuscript. The manuscript was written by Zachary Kelly with equal contributions from both authors regarding its contents and direction. \\


\noindent
{\large \textbf{Acknowledgements}} \\

\noindent
The authors recognize the support of this project from Academics Without Borders. They would also like to give special thanks to the University of Eswatini, in particular Simiso K. Mkhonta and Sandile Motsa, for their support and involvement in this project and proposal, and to Joseph Menezes, Harshil Vyas and Philip Nadon for their contributions in the early phase of the project. \\

\end{document}